\documentclass[conference]{IEEEtran}

\usepackage{booktabs} 

\usepackage{graphicx}

	\usepackage{amsmath}
	\usepackage{amsfonts}
    \usepackage{longtable}
	\usepackage{fancyhdr}
	\usepackage{mdframed}
\mdfdefinestyle{style1}{leftmargin=1cm,rightmargin=0.5cm,skipbelow=0.5cm,skipabove=0.2cm,
   innertopmargin=2pt}
	\usepackage{balance}
    \usepackage{fancybox}
    \usepackage{booktabs}
	\usepackage{enumerate}
    \usepackage{framed}
	\usepackage{listings}
    \usepackage{subcaption}
    \usepackage{xcolor}
    \usepackage{hyperref}

\ifCLASSOPTIONcompsoc
  \usepackage[nocompress]{cite}
\else
  \usepackage{cite}
\fi

\begin{document}
\title{Towards Human-Like Automated Test Generation: Perspectives from Cognition and Problem Solving}


\author{Eduard Enoiu$^{1}$, Robert Feldt$^{2}$
\\$^{1}$School of Innovation, Design and Engineering, M\"alardalen University, V\"aster\aa s, Sweden%
\\$^{2}$Department of Computer Science and Engineering, Chalmers University of Technology, Gothenburg, Sweden%
}

\IEEEtitleabstractindextext{%
\begin{abstract}
Automated testing tools typically create test cases that are different from what human testers create.
This often makes the tools less effective, the created tests harder to understand, and thus results in tools providing less support to human testers.
Here, we propose a framework based on cognitive science and, in particular, an analysis of approaches to problem solving, for identifying cognitive processes of testers.
The framework helps map test design steps and criteria used in human test activities and thus to better understand how effective human testers perform their tasks.
Ultimately, our goal is to be able to mimic how humans create test cases and thus to design more human-like automated test generation systems. We posit that such systems can better augment and support testers in a way that is meaningful to them.







\end{abstract}
}
%
%



\maketitle
\IEEEdisplaynontitleabstractindextext
\IEEEpeerreviewmaketitle

\section{Introduction}

The creation of test cases in software testing is an intellectual activity in which engineers allocate a variety of cognitive resources when confronted with challenges, as they go along. This is largely a manual activity, dependent on the ingenuity and thoroughness of humans. Automated test generation \cite{anand2013orchestrated} has been proposed to allow test cases to be created with less effort.
The goal is to automatically find a small set of test cases that check the correctness of the system and guard against (previous as well as future) faults. While a lot of progress has been made, it remains a challenge to create strong as well as small test suites that are also relevant to developers. Consequently, industry still mainly rely on manual testing.

Given the emerging evidence (e.g., \cite{fraser2015does,enoiu2016controlled}) suggesting that \textit{automated test generation cannot match the fault detection effectiveness of manually created tests}, the purpose herein is to propose a framework to help understand and, ultimately, to mimic\slash simulate the problem-solving behaviors of testers. 
Recently, Gay et al. \cite{gay2017fitness,gay2019one} proposed some strategies for improving automated test data generation using adaptation and context-based characteristics. However, context and domain-specific information is just some of the dimensions used by human testers. 
While many of these test generation approaches are AI\slash Machine Learning (ML) oriented, we are concerned with grounding this framework in high-level cognitive theories and psychological data. For example, we should certainly not treat search-based test generation as a theory of human-based testing, in the same way as AlphaGo is not used as a theory of human Go.
While this shows that automated systems can beat humans, we argue that when they do not we have something to gain from modeling human cognition and problem solving and, potentially, mimicking it in our tools.

 \begin{figure*}[ht!]
  	\centering   
    \includegraphics[width=0.70\textwidth]{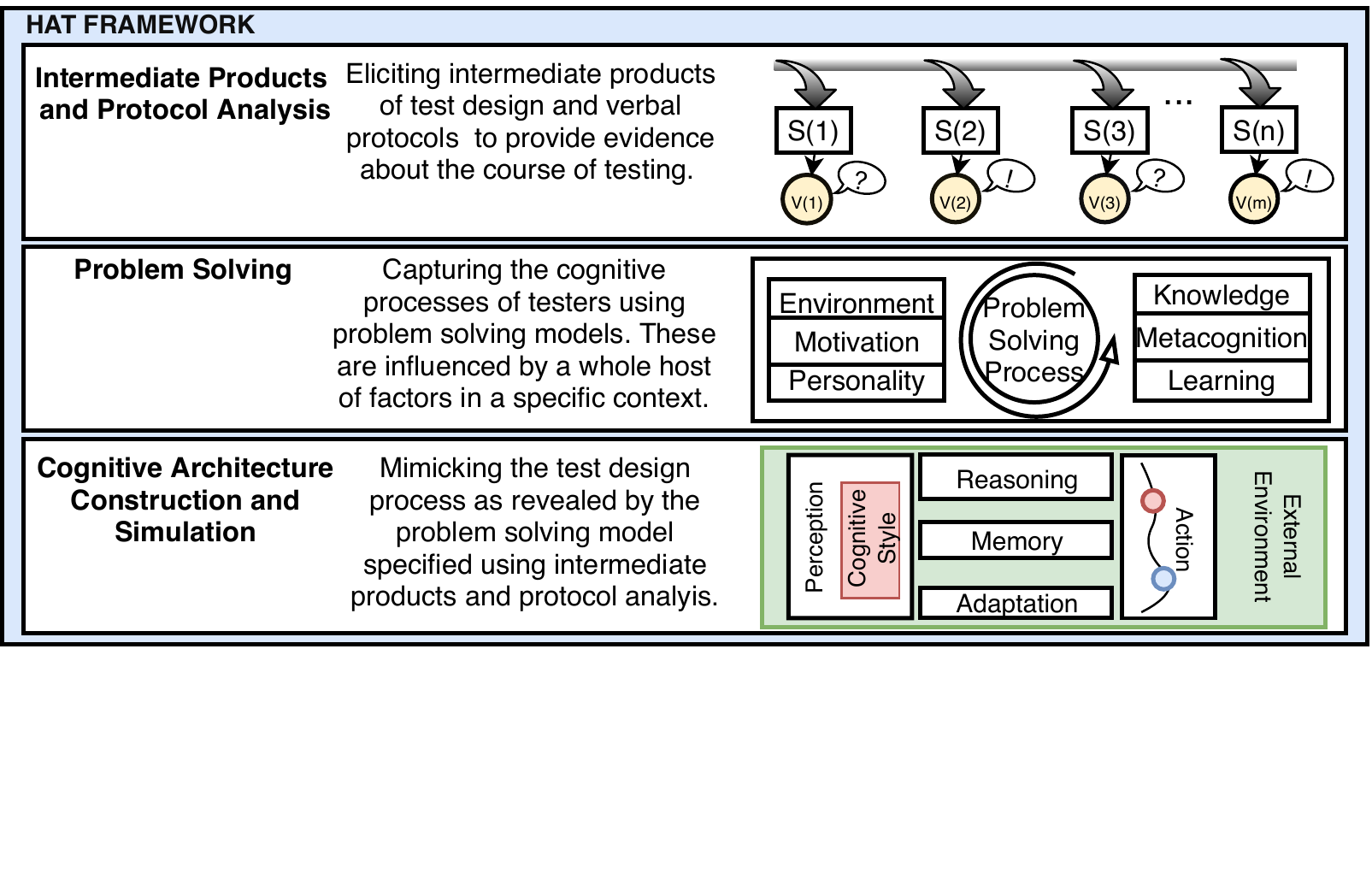}
    \vspace{-20mm}
 	\caption{Outline of the HAT framework and its key features.}
 	\label{fig:model}
\end{figure*}

\section{The HAT Framework}
The \textit{Human-based Automated Test (HAT)} generation framework assembles cognitive processes used by testers into a problem solving model and cognitive architecture that can be used for improving automated test generation\footnote{This work is partially supported by Software Center, EU's Horizon 2020 program (No. 957212) and by Vinnova through the XIVT project.}. Rather than directly trying to build more `intelligent' automated test generation systems, we outline a framework for the study of the human mind when creating test cases. 
Figure \ref{fig:model} provides a high level summary of the key components of the framework. It uses a problem solving approach\cite{newel1972human,pretz2003recognizing} to depict the behavior by which testers perform test creation activities. The HAT framework proposes the use of cognitive architectures and simulation to discover the conditions when humans create effective test cases. 
Below we outline its three main levels.

\subsubsection{Intermediate Products and Protocol Analysis}
If we are interested in how people create test cases, we need to collect information about the various cognitive actions and decisions taken during the process. Whereas the created test cases typically do not convey their underlying creation process, this may be revealed by studying the intermediate products \cite{jaarsveld2005sketches}. Testers often create sketches or note useful information in the course of creating test cases. These intermediate products provide much finer constraints (states S(x) of a cognitive process in Figure \ref{fig:model}) on possible explanations of certain behaviors. These can be complemented by verbal protocol \cite{ericsson1998protocol} analysis using a think aloud method that provides information about the course of test creation (verbal reports v(y) in Figure \ref{fig:model}). Although protocol analysis is quite old, and it has its limitations, there is evidence \cite{hughes2003trends,itkonen2012role} supporting the use of verbal reports for providing valid data about what humans are thinking and serve as the basis for theories of testing. 

\subsubsection{Problem Solving}
It is clear that humans involved in software testing are basically trying to solve a sequence of test problems\slash challenges. Therefore, examining test case design as a problem solving process can help us explain how individuals are able to apply their knowledge to existing and novel problems. The results of analyzing intermediate products and verbal protocols should be used to provide finer constraints and possible accounts of how people solve the problem of creating test cases. Problem solving is a very important topic of research in cognitive science but also in applied fields such as artificial intelligence. 
In software engineering, the cognitive models developed by Hale et al.\cite{hale1999evaluation} and Enoiu et al. \cite{enoiu2020towards} and instantiated for the specific problem of creating test cases and debugging seem to be the closest models to the HAT framework for capturing the cognitive processes of testers. Nevertheless, since the classical models these results were based on are quite limited, they should be complemented with other cognitive theories that can handle different psychological issues of human-based test creation. 
However, problem solving is not enough and many factors affect humans during testing, e.g. motivation, creativity, emotion, transfer of knowledge, language parsing, expertise, etc.
Over time, we will thus need to elaborate the framework to cater also to other psychological as well as, eventually, social factors.

\subsubsection{Cognitive Architecture Construction and Simulation}
Problem solving in software testing, learning and the development of testing expertise are behaviours that require rigorous explanations. Just as a software architect may produce an architectural model that shows how different parts of the software are working together, so are we interested in building a model that incorporates a theory of how testers think and learn. One way to mimic the captured problem solving process is to build a general model of architecture in which a variety of cognitive processes can be specified as runnable computational models (e.g., algorithms guiding the test generation). Simulations are then performed to check if the posited processes are sufficient to lead to the observed protocol behavior. While there is no cognitive architecture for software testing, several general architectural models of cognition exist \cite{kotseruba202040} such as ACT-R, SOAR and CLARION. Such cognitive models could help us reverse engineer the human tester--- the only available intelligent test design system. These cognitive architectures could constitute a solid basis for developing human-like automated test generation tools, properly grounded in existing cognitive research.

\section{Conclusions}
We conclude that the HAT framework is a basis for employing intermediate products and protocol analysis to build problem solving models and cognitive architectures that can be used in the end for advancing the understanding of the mind of humans performing software testing. Therefore, we argue that this framework can help
enhance automated test generation as well to better accommodate the cognitive needs of human testers with the testing tools.

\balance

\bibliographystyle{IEEEtran}
\bibliography{acmart} 

\end{document}